\documentclass[fleqn,usenatbib]{mnras}

\usepackage{newtxtext,newtxmath}

\usepackage[T1]{fontenc}

\DeclareRobustCommand{\VAN}[3]{#2}
\let\VANthebibliography\thebibliography
\def\thebibliography{\DeclareRobustCommand{\VAN}[3]{##3}\VANthebibliography}

\usepackage{graphicx}	
\usepackage{amsmath}	

\title[Age dependence of type Ia SN luminosity]{Evidence for strong progenitor age dependence of type Ia supernova luminosity standardization process}

\author[Y. -W. Lee et al.]{
Young-Wook Lee,$^{1}$\thanks{E-mail: ywlee2@yonsei.ac.kr (YWL)}
Chul Chung,$^{1}$\thanks{E-mail: chulchung@yonsei.ac.kr (CC)}
Pierre Demarque,$^{2}$
Seunghyun Park,$^{1}$
Junhyuk Son$^{1}$
and Yijung Kang$^{3}$
\\
$^{1}$Department of Astronomy \& Center for Galaxy Evolution Research, Yonsei University, Seoul 03722, Republic of Korea\\
$^{2}$Department of Astronomy, Yale University, New Haven, CT 06520-8101, USA\\
$^{3}$Gemini Observatory/NSF’s NOIRLab, Casilla 603, La Serena, Chile
}

\date{Accepted XXX. Received YYY; in original form ZZZ}

\pubyear{2022}
\defcitealias{1956PASP...68....5B}{Baade's (1956)}
\setcitestyle{notesep={; },round,aysep={},yysep={;}}

\begin{document}
\label{firstpage}
\pagerange{\pageref{firstpage}--\pageref{lastpage}}
\maketitle

\begin{abstract}
Supernova (SN) cosmology is based on the assumption that the width-luminosity relation (WLR) and the color-luminosity relation (CLR) in the type Ia SN luminosity standardization would not show {absolute magnitude differences} with progenitor age.
Unlike this expectation, recent age datings of stellar populations in host galaxies have shown significant correlations between progenitor age and Hubble residual (HR).
Here we show that this correlation originates from a strong progenitor age dependence of {the zero-points of} the WLR and the CLR, in the sense that SNe from younger progenitors are fainter each at given light-curve parameters $x_1$ and $c$.
This $4.6\sigma$ result is reminiscent of Baade's discovery of {the zero-point variation of the} Cepheid period-luminosity relation {with age}, and, as such, causes a serious systematic bias with redshift in SN cosmology.
Other host properties show substantially smaller and insignificant {offsets} in the WLR and CLR for the same dataset.
We illustrate that the differences between the high-$z$ and low-$z$ SNe in the WLR and CLR, and in HR after the standardization, are fully comparable to those between the correspondingly young and old SNe at intermediate redshift, indicating that the observed dimming of SNe with redshift may well be an artifact of over-correction in the luminosity standardization.
When this systematic bias with redshift is properly taken into account, there is little evidence left for an accelerating universe, {in discordance with other probes,} urging the follow-up investigations with larger samples at different redshift bins.
\end{abstract}




\section{Introduction}
\label{s1}

Supernova (SN) cosmology has long been considered to provide {the most direct evidence for an accelerating universe.}
Based solely upon the Hubble diagram, it is a very simple method and is less dependent on the model compared to other cosmological probes \citep{2008ARA&A..46..385F, 2013PhR...530...87W}.
Cosmic microwave background (CMB) provides crucial cosmological constraints, notably the geometry of the universe, but ``it alone provides relatively weak constraints on dark energy'' \citep{2020A&A...641A...6P}.
Therefore, SN cosmology is {arguably} best suited for directly investigating an accelerating universe.
More than two decades ago, \citet{1998AJ....116.1009R} and \citet{1999ApJ...517..565P} discovered that SNe at high redshift are fainter by 0.20 - 0.25~mag ($\sim$20\% in brightness) compared to the model without the dark energy, after the empirical SN luminosity standardization.
This was interpreted as evidence {for an accelerating universe, which is currently considered strongly supported by other lines of evidence \citep{2020A&A...641A...6P, 2021PhRvD.103h3533A} from CMB and baryon acoustic oscillations (BAO).}

Although this interpretation from SNe is widely accepted, an alternative interpretation is possible for the observed dimming of SNe with redshift.
As first pointed out by \citet{1968ApJ...151..547T}, the luminosity evolution of standard candle with redshift can be a serious systematic error in observational cosmology, and several investigators suggested that this possibility should be considered in detail in SN cosmology {as well} \citep{2000ApJ...530..593D, 2009A&A...506.1095L, 2017A&A...602A..73T}.
Within the redshift range most relevant to SN cosmology ($0 < z < 1$), we expect {a significant variation in the SN progenitor age distribution \citep{2014MNRAS.445.1898C}. 
Therefore, we cannot rule out the possibility that SNe at high redshift are fainter, not because of an accelerating universe, but because their progenitors are younger, in the mean, compared to their counterparts at the local universe.}

\begin{figure*}
\centering
\includegraphics[angle=0,scale=0.32]{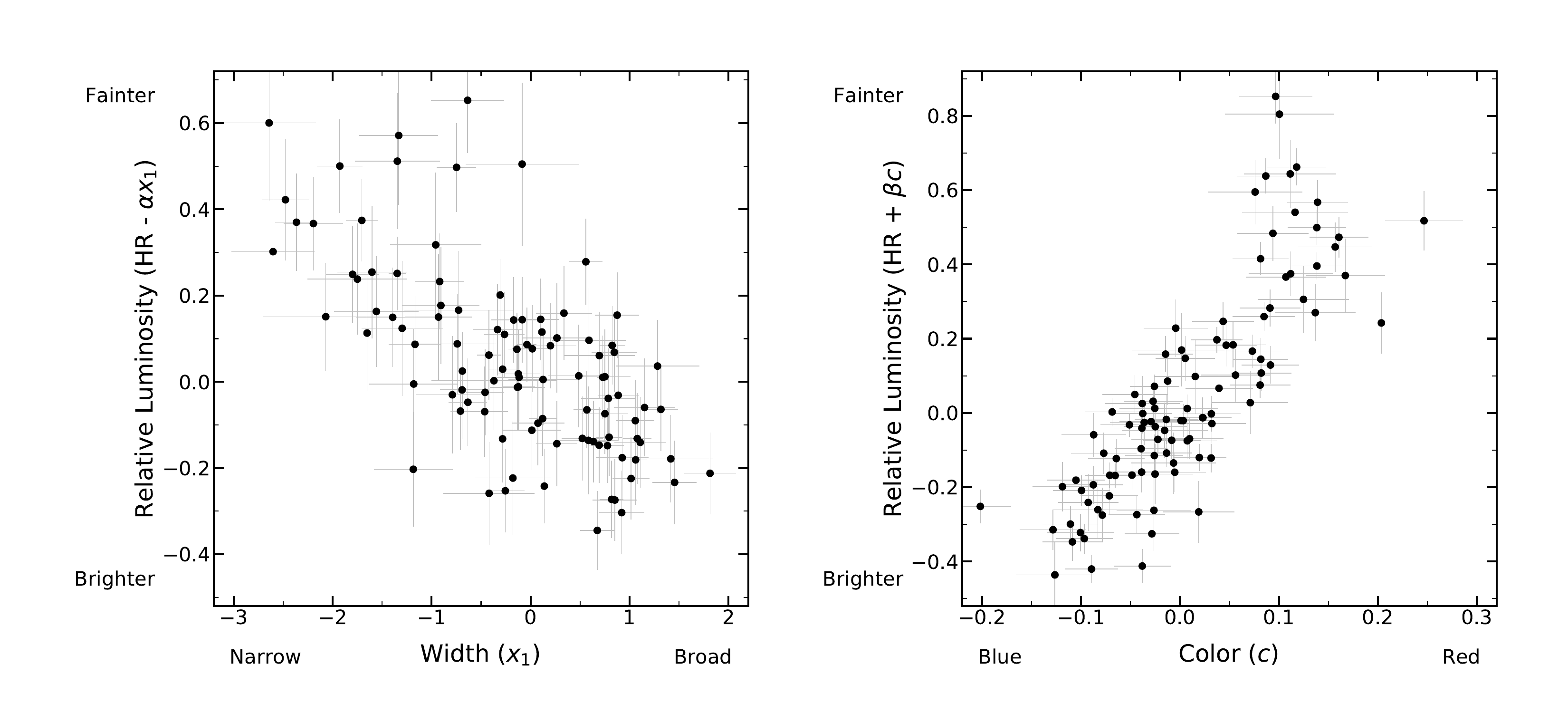}
\caption{ 
The width - luminosity relation (WLR) and color - luminosity relation (CLR) for 102 SNe Ia in \citet{2019ApJ...874...32R} sample at $0.05 < z < 0.20$ with a median $z = 0.14$.
Like the Cepheid period-luminosity relation, the SN luminosity standardization is based on the WLR and CLR. 
Following \citet{2006A&A...447...31A}, the left panel computes distance modulus $\mu_{\rm SN}$ without the width term $\alpha x_1$ (corrected only for $c$), while the right panel computes $\mu_{\rm SN}$ without the color term $\beta c$ (corrected only for $x_1$), to recover WLR and CLR, respectively. 
As adopted and suggested by \citet{2019ApJ...874...32R}, the light curve data ($x_0$, $x_1$, \& $c$) are from \citet{2013ApJ...763...88C}, and the HR's are calculated with $\alpha = 0.16$, $\beta = 3.12$, and $M_x = -29.65$ ($M_B = -19.01$) from the $\Lambda$CDM baseline model ($h = 0.738$, $\Omega_{M} = 0.24$, $\Omega_{\Lambda} = 0.76$).
\label{f1}}
\end{figure*}

As is well known, \citet{1956PASP...68....5B} realized that young population~I Cepheids that Hubble discovered in M31 are in fact brighter, at a given period, than old population II counterparts based on which Hubble calibrated his observations. 
Because of this discovery, distances to M31 and other galaxies have been doubled, and the value of the Hubble parameter has been decreased by a factor of two. 
This illustrates that the luminosity of a standard candle can depend on stellar population age because of the difference in stellar mass. 
If a similar shift with progenitor age is discovered among SNe Ia, this would also have a critical impact on cosmology because SN cosmology is based on the assumption that the width-luminosity and color-luminosity relations in the type Ia supernova (SN Ia) luminosity standardization would not show {absolute magnitude differences} with progenitor age and redshift \citep{2019NatAs...3..706J}. 
More than a decade ago, \citet{2010MNRAS.406..782S}, in their pioneering investigations, found that SN Ia luminosities, after standardization, depend on the global properties of host galaxies, such as host mass and specific star formation rate. \citet{2010ApJ...715..743K} also reached at the same conclusion that the Hubble residuals (HRs) of SNe Ia are correlated with host galaxy masses \citep[see also][]{2013ApJ...770..108C, 2013MNRAS.435.1680J}. 
More recent investigations by \citet{2013A&A...560A..66R, 2020A&A...644A.176R} further showed a strong dependence on the local specific star formation rate (LsSFR) of host galaxies \citep[see also][]{2022A&A...657A..22B}. 
Since the host mass and the physics of star formation cannot directly affect the luminosity of a SN in a host galaxy, these correlations are presumably not directly originated from host mass and LsSFR, but more likely due to progenitor age or metallicity closely related to these properties \citep{2010MNRAS.406..782S, 2016ApJS..223....7K, 2020A&A...644A.176R}.

\begin{figure*}
\centering
\includegraphics[angle=0,scale=0.32]{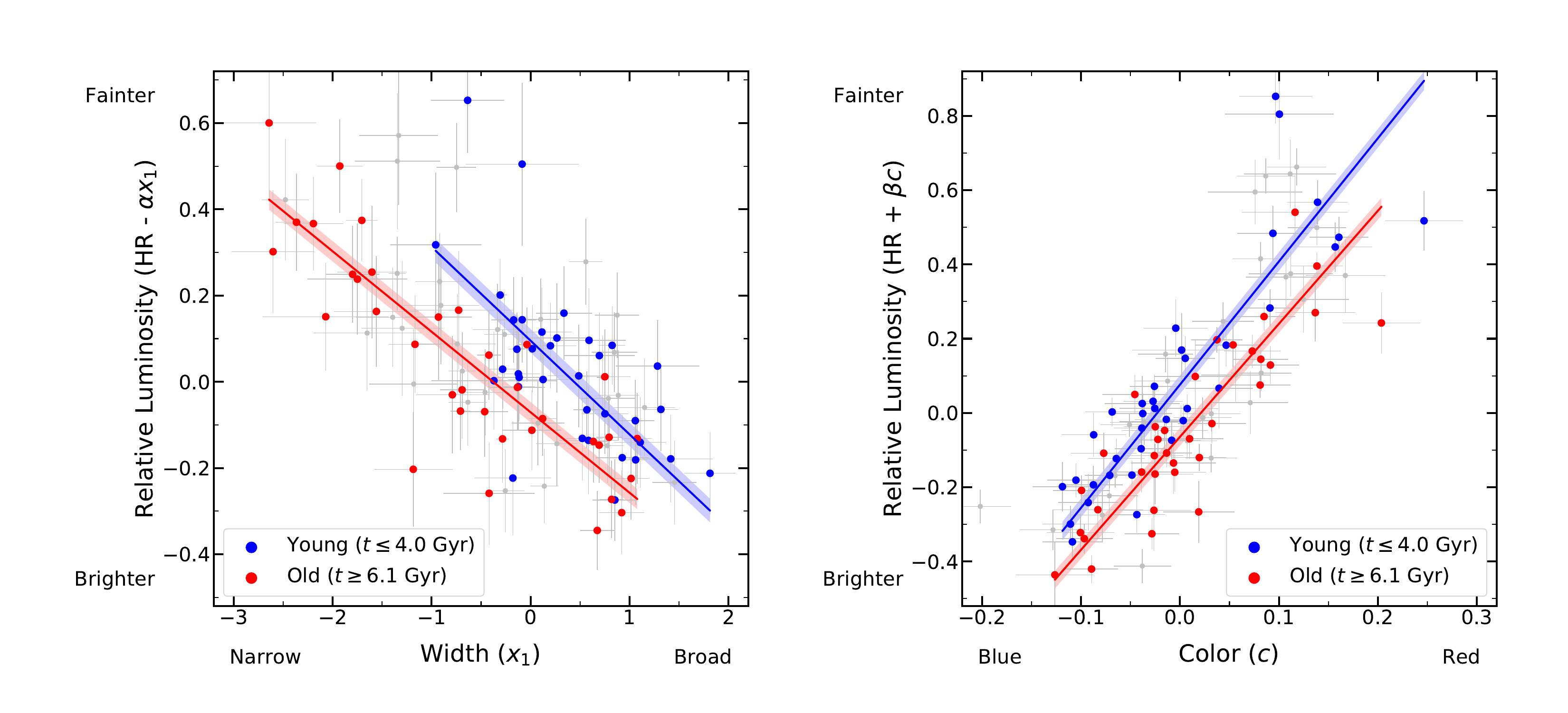}
\caption{
Same as {Figure~\ref{f1}}, but the sample is subdivided into two according to the population age. 
SNe from young and old populations have nearly the same sample size ($N = 36$ \& $35$) with a gray zone between.  
Strong progenitor age dependence is discovered in the sense that SNe from younger progenitors are fainter each at given light-curve parameters $x_1$ and $c$. 
This is reminiscent of \citetalias{1956PASP...68....5B} discovery of two Cepheid period-luminosity relations, and, as such, should be universal and have a critical impact on SN cosmology.
The blue and red lines are the regression fits (with $\pm 1 \sigma$ intercept error) from MCMC posterior sampling method for young and old progenitors, respectively.
When measured at $x_1 = 0.0$, the difference between the two age subgroups is $0.166 \pm 0.036$~mag (i.e., $\Delta {\rm HR}/ \Delta {\rm age} = -0.040$~mag/Gyr).
\label{f2}}
\end{figure*}

{In order to directly measure reliable population ages for host galaxies, \citet{2020ApJ...889....8K} obtained very high quality spectra for 59 nearby early-type host galaxies (ETGs).}
Excluding some abnormal ETGs with recent star formation (SF), {they} found an important $\sim$$3 \sigma$ correlation between population age and HR, in the sense that SNe in younger host galaxies are fainter, which would indicate a luminosity evolution in SN cosmology because high-$z$ SNe should be from younger progenitors.
\citet{2020ApJ...896L...4R}, however, claimed that this result from ETGs is not confirmed from a larger sample of host galaxies comprising all morphological types.
Their claim was based on the two age datasets, one by \citet{2018ApJ...867..108J} and the other by \citet{2019ApJ...874...32R}, all measured from multi-band optical photometry.
In our rebuttal paper \citep{2020ApJ...903...22L}, we have shown that the \citet{2018ApJ...867..108J} ages, in particular, are based on highly uncertain and inappropriate luminosity-weighted ages derived, in many cases, under serious template mismatch.
The other dataset is based on the improved photometric age dating of \citet{2019ApJ...874...32R} for mass-weighted ages implemented in the updated version of the population synthesis model of \citet{2010ApJ...712..833C}, which, unlike luminosity-weighted ages, are not biased by on-going or recent SF.
We found, however, that the statistical analysis of \citet{2020ApJ...896L...4R} is seriously affected by the regression dilution bias, severely underestimating both the slope and significance of the age-HR correlation.
When the regression analysis is performed with an MCMC posterior sampling method \citep{2007ApJ...665.1489K}, a very significant (4.3$\sigma$) correlation is obtained between population age and HR for both global age and local age near the site of SN with the slope in excellent agreement with our previous spectroscopic result from ETGs \citep{2020ApJ...889....8K}.
Recently, \citet{2021MNRAS.503L..33Z} confirms that this age-HR correlation is statistically significant ($5 \sigma$), although they obtained a somewhat shallower slope from a posterior sampling method adopting full posterior for the age error instead of the Gaussian error.
Therefore, even the dataset originally used by \citet{2020ApJ...896L...4R} to oppose our claim is instead strongly supporting our result, and the luminosity evolution stands up to scrutiny as a serious systematic bias in SN cosmology.
It was not clear, however, how this correlation arises from the SN luminosity standardization process, and how this would impact the cosmological result.
The same dataset of \citet{2019ApJ...874...32R} for reliable mass-weighted ages, with a sufficiently large sample size (N = 102) coupled with adequate age accuracy, makes it possible to look into the SN luminosity standardization process in a more detailed manner.
This dataset for directly measured population ages near the SN sites in host galaxies comprising all morphological types is currently best suited for such analysis.

\begin{figure*}
\centering
\includegraphics[angle=-90,scale=0.65]{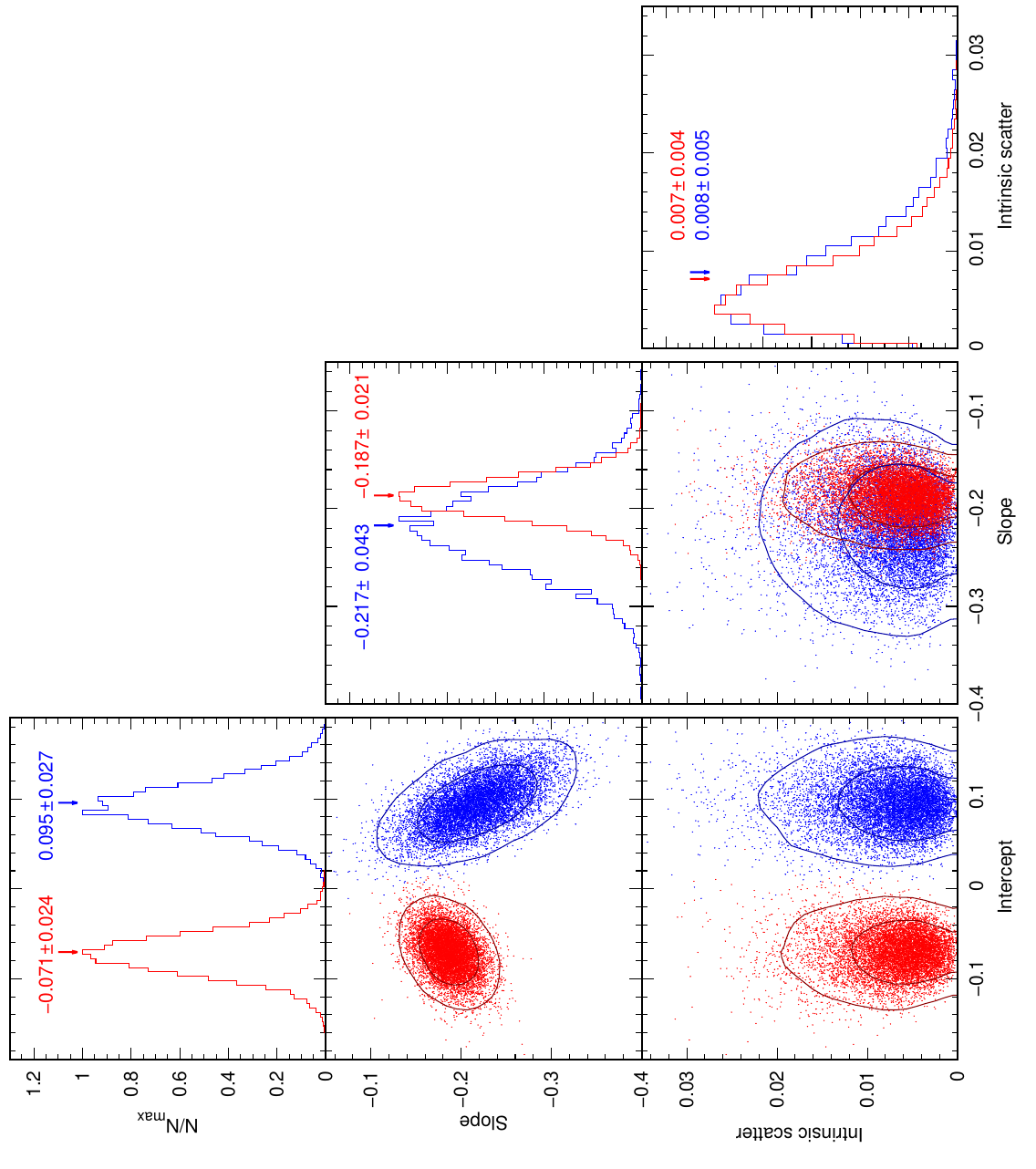}
\caption{
Corner plot of the posterior distributions for the parameters of the regression models for the young and old subgroups (the blue and red histograms, respectively) in the left panel of {Figure~\ref{f2}}.
The mean values of the distributions are indicated with arrows.
\label{f3}}
\end{figure*}

\begin{figure*}
\centering
\includegraphics[angle=0,scale=0.32]{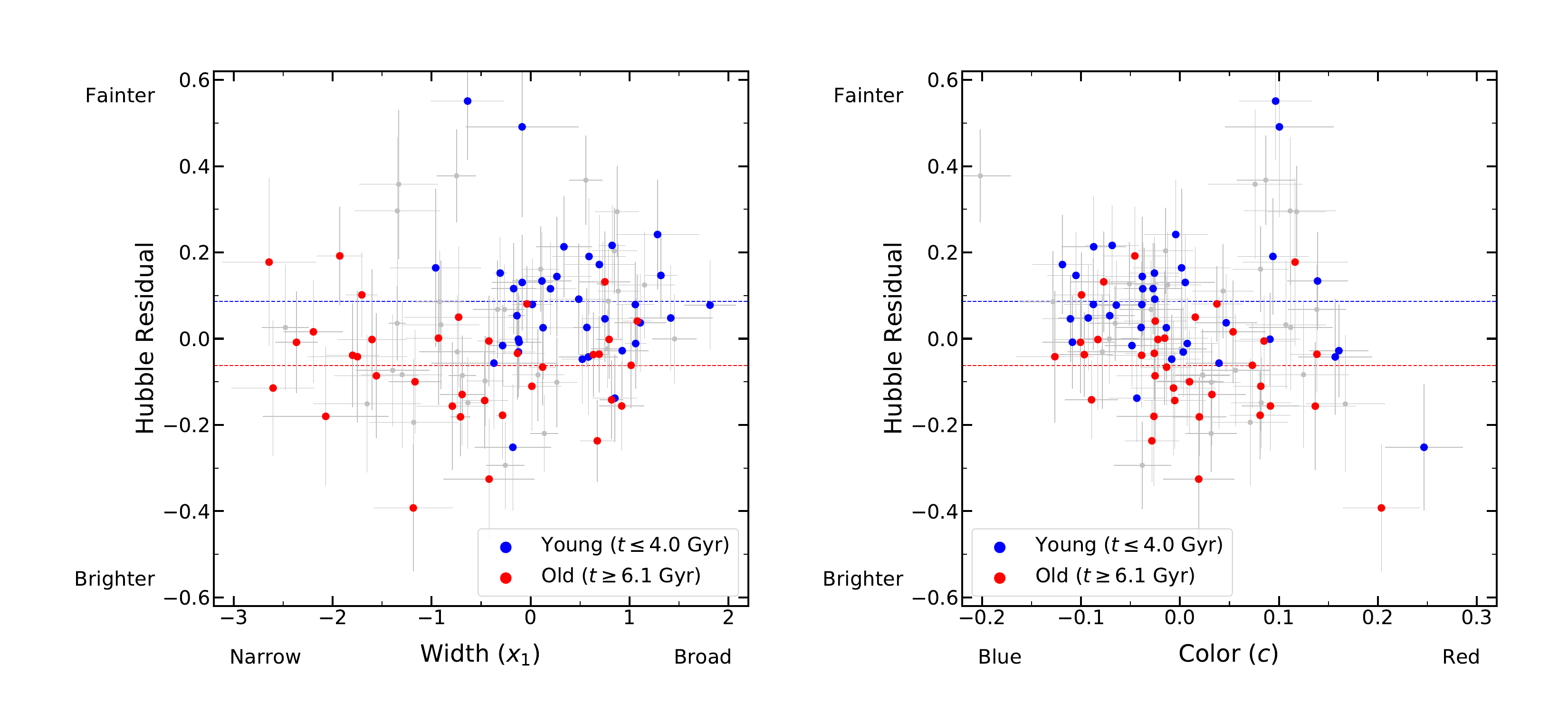}
\caption{
Similar to {Figure~\ref{f2}}, but after the luminosity standardization, which is a rotation of the WLR (or CLR) in {Figure~\ref{f2}} according to the slope $\alpha$ (or $\beta$).
The dashed lines are the mean values of HR.
Note that SNe from younger progenitors are over-corrected and become relatively fainter.
SNe at high redshift are also from the younger population, and, therefore, should be equally over-corrected and become similarly fainter.
\label{f4}}
\end{figure*}

\section{Progenitor age dependence of the width-luminosity and color-luminosity relations}
\label{s2}

Type Ia SN luminosity standardization process is based on the light curve width (stretch) parameter ($\Delta m_{15}$, $s$, or $x_1$) together with color parameter $c$ \citep{1993ApJ...413L.105P, 1998A&A...331..815T}.
In the modern SALT2 method \citep{2007A&A...466...11G}, the distance modulus from SN Ia is given by
\begin{equation}
\mu_{\rm SN} = m_B - M_B + \alpha x_1 - \beta c ,
\label{e1}
\end{equation}
where $m_B$ is an apparent magnitude, $M_B$ is an absolute magnitude at the reference point ($x_1 = 0.0$, $c = 0.0$), and $\alpha x_1$ and $\beta c$ are the correction terms depending on the light curve width ($x_1$) and color ($c$).
Therefore, like the Cepheid period-luminosity relation, the SN luminosity standardization relies on the width-luminosity relation (WLR) and the color-luminosity relation (CLR), in which the absolute values of the slopes of the correlations are $\alpha$ and $\beta$, respectively.
These relations are also called ``brighter-broader'' and ``brighter-bluer'' relations.
The key assumption of SN cosmology is that these relations should not depend on progenitor age, because high-$z$ SNe should be from relatively younger progenitors.
This is well described in \citet{2019NatAs...3..706J}, ``if SNe Ia are to be good standardizable candles over cosmic time, the calibrating  relationships between SN luminosity and light-curve shape must be invariant with progenitor age''.
Our finding of the correlation between population age and HR raises a question as to the validity of this key assumption in SN cosmology.
In order to understand this, we have investigated below how the population age affects the WLR and CLR of the SN luminosity standardization process.

\begin{figure*}
\centering
\includegraphics[angle=0,scale=0.22]{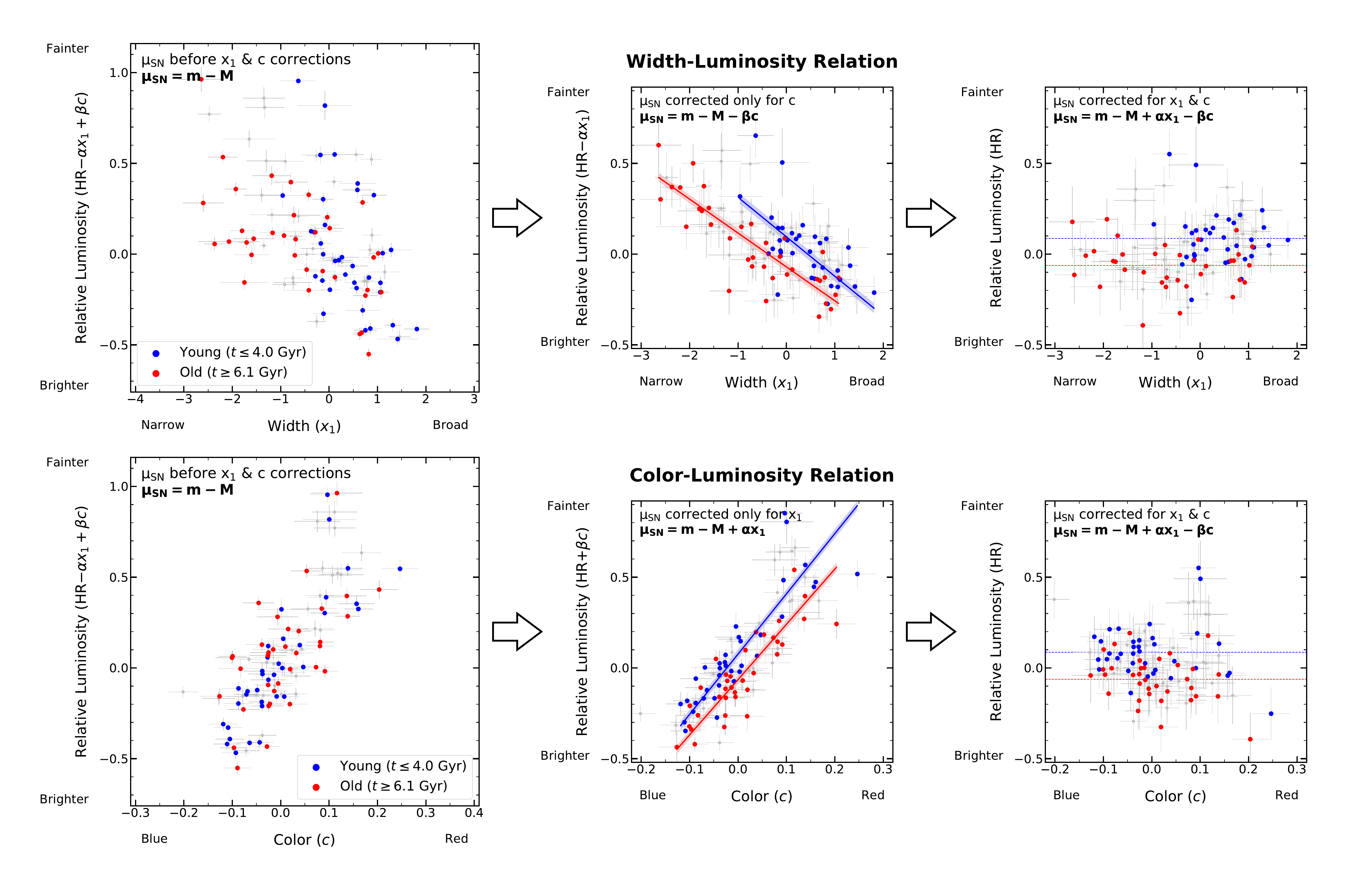}
\caption{
The whole process of SN Ia luminosity standardization is illustrated on the WLR (top panels) and CLR (bottom panels), respectively.
The middle and right panels are the same as {Figures~\ref{f2} and \ref{f4}}, but the left panels are added to show the WLR and CLR before $x_1$ and $c$ corrections.
When the $\mu_{\rm SN}$ is corrected for $c$ ($x_1$), the WLR (CLR) shows a more clear split between the young and old progenitors (middle panels).
After the luminosity standardization ($\mu_{\rm SN}$ corrected for both $x_1$ and $c$), SNe from younger progenitors are over-corrected and become relatively fainter (right panels).
\label{f5}}
\end{figure*}

When the SN sample is confined to a narrow redshift range, as in the \citet{2019ApJ...874...32R} sample, the relative luminosities of SNe Ia can be compared to each other from the values of the HR, which is defined by
\begin{equation}
{\rm HR} = \mu_{\rm SN} - \mu_{model},
\label{e2}
\end{equation}
where $\mu_{model}$ is the distance modulus at the redshift $z$ according to an assumed cosmological model.
In {Figure~\ref{f1}}, we plot the WLR and CLR for the SNe Ia in \citet{2019ApJ...874...32R} sample, which is a subset of the SDSS SN survey \citep{2013ApJ...763...88C} at $0.05 < z < 0.20$ with a median $z = 0.14$.
 Following \citet{2006A&A...447...31A}, the left panel computes distance modulus $\mu_{\rm SN}$ without the width term $\alpha x_1$ (corrected only for $c$), while the right panel computes $\mu_{\rm SN}$ without the color term $\beta c$ (corrected only for $x_1$), to recover WLR and CLR, respectively.
As adopted or suggested by \citet{2019ApJ...874...32R}, the light curve data ($x_0$, $x_1$, \& $c$) are from \citet{2013ApJ...763...88C}, and the HR's are calculated with $\alpha = 0.16$, $\beta = 3.12$, and $M_x = -29.65$ ($M_B = -19.01$) from the $\Lambda$CDM baseline model ($h = 0.738$, $\Omega_{M} = 0.24$, $\Omega_{\Lambda} = 0.76$).
It is clear that the \citet{2019ApJ...874...32R} sample well follows the ``brighter-broader'' and ``brighter-bluer'' relations.

To investigate the progenitor age dependence, in {Figure~\ref{f2}}, we subdivide this sample into two according to the population age with a gray zone between, so that SNe from younger (age $\leq 4$~Gyr) and older  (age $\geq 6.1$~Gyr) populations have nearly the same sample size ($N = 36$) with  31 SNe in a gray zone.
The split between the {``young''} and {``old''} subgroups was chosen so that the width of the gray zone equals (or be larger than) the mean measurement error (2.0~Gyr) to avoid sample contamination between the two subgroups while maintaining a sufficient sample size ($N > N_{\rm total}/3 = 34$) for each subgroup.
As described above, the population ages of host galaxies adopted for this analysis are from the improved and reliable photometric age dating for mass-weighted ages.
\citet{2019ApJ...874...32R} measured both the global ages of host galaxies and the local ages near the sites of SNe, and we employed the local ages because they should be more relevant to the SN progenitor ages.
The blue and red lines are the regression fits (with $\pm 1 \sigma$ intercept error) from the MCMC posterior sampling method (implemented in the LINMIX package of \citealt{2007ApJ...665.1489K}) for the young and old progenitors, respectively.
{Obviously}, there is a strong population age dependence in the sense that SNe from younger progenitors are fainter each at given $x_1$ and $c$.
Therefore, the zero-points of the WLR and CLR {vary significantly with age ($\Delta{\rm mag} = 0.166 \pm 0.036$), while the slopes of these relations appear to be little affected ($\Delta {\alpha} = 0.031 \pm 0.047$; $\Delta {\beta} = 0.27 \pm 0.51$).
This is qualitatively consistent with early findings from host mass and LsSFR (see Section~\ref{s4}).}
When measured at $x_1 = 0.0$, the difference between the two age subgroups is $0.166 \pm 0.036$~mag for the population age difference of 4.2~Gyr (i.e., $\Delta {\rm HR}/ \Delta {\rm age} = -0.040$~mag/Gyr).\footnote{This magnitude offset is comparable to the LsSFR step of \citet{2020A&A...644A.176R}, who have interpreted their result as a progenitor age effect.
Like their LsSFR step, our magnitude offset is quite higher than the rest of the literature (see Section~\ref{s4} for the comparison with the literature).}
This result is significant at $4.6 \sigma$ level as estimated from the intercept errors of two regression fits (see {Figure~\ref{f3}} for posterior distributions).\footnote{ To estimate the confidence level, we fit the posterior distribution with a Gaussian function and computed the standard deviation for each subpopulation.
With these two values for the intercept errors, together with the intrinsic scatters, the uncertainty for the difference was calculated from the propagation of errors.}
We have also performed the Bonferroni correction to account for multiple testing, which shows that the significance of our result is $4.2 \sigma$ even in this conservative test.
The population ages can be roughly converted to the SN progenitor ages based on Figure~3 of \citet{2014MNRAS.445.1898C}, according to which the difference in progenitor age would correspond to $\sim$3.0~Gyr between the two subgroups (i.e.,  $\Delta {\rm HR}/ \Delta {\rm age} = -0.056$~mag/Gyr).

\begin{figure*}
\centering
\includegraphics[angle=0,scale=0.64]{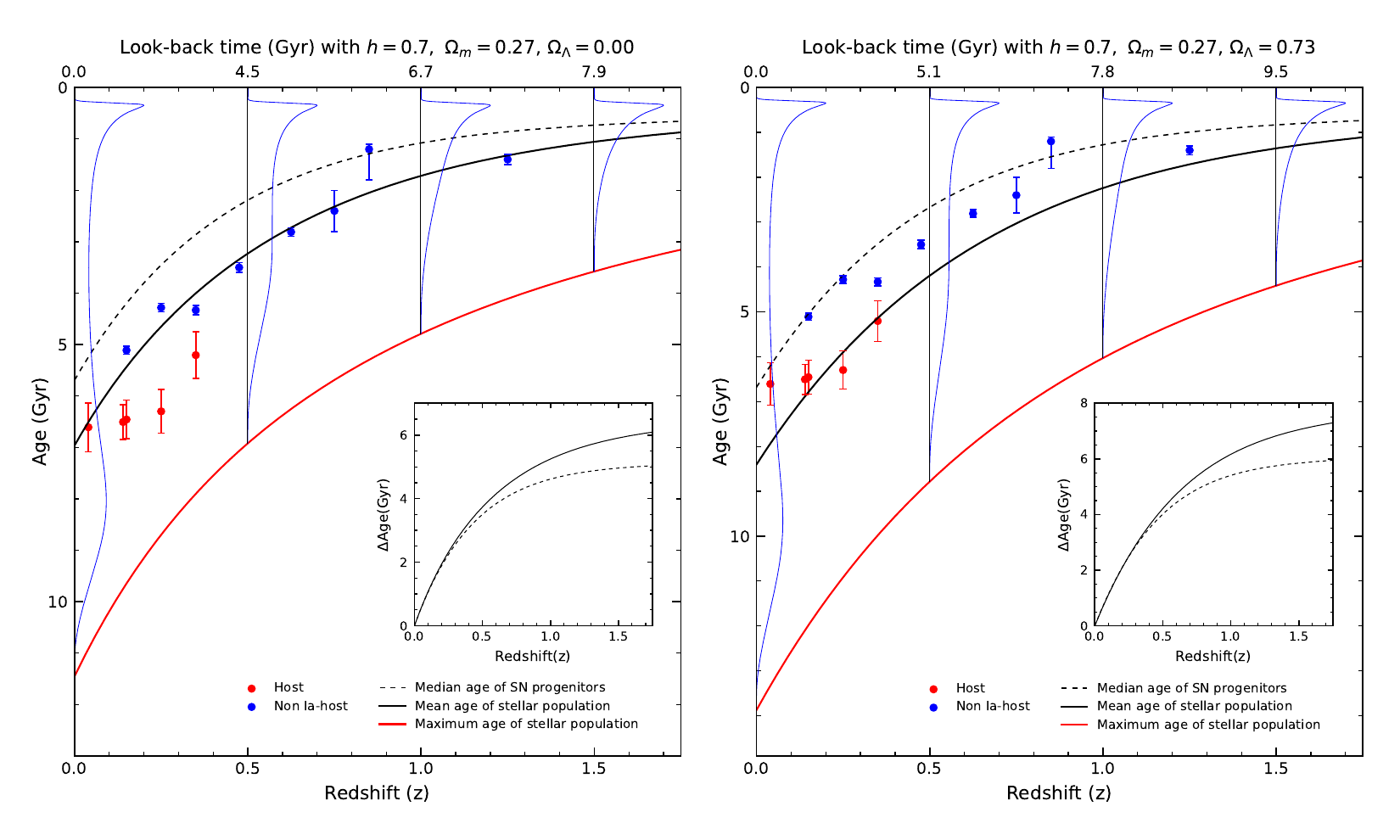}
\caption{ 
Evolution of stellar population age with redshift under two different cosmological models ($\Omega_{\Lambda} = 0.00$ \& $\Omega_{\Lambda} = 0.73$).
 Stellar populations and SN progenitors in host galaxies get younger with redshift.  
The distribution functions (the blue lines) at $z = 0.0$, 0.5, 1.0, \& 1.5 are SPADs derived in {Figure~\ref{f7}}. 
The black dashed line is for the median age of SN progenitors, while the black solid line is for the mass-weighted mean age of stellar population obtained from cosmic star formation history.
The red line is for the maximum age of stellar population.
The inset is for the difference with respect to $z = 0$. 
Observed data compared are galaxy mass-weighted average values of mean population ages of host \citep{2011ApJ...740...92G, 2019ApJ...874...32R, 2020ApJ...889....8K} and {presently} {non~Ia-host} \citep{2006ApJ...651L..93S, 2014ApJ...792...95C, 2016ApJ...822....1F} galaxies of comparable stellar mass ($\log M/M_{\odot} = 10.3$ -- $11.3$) at different redshifts.
\label{f6}}
\end{figure*}

\begin{figure*}
\centering
\includegraphics[angle=0,scale=0.59]{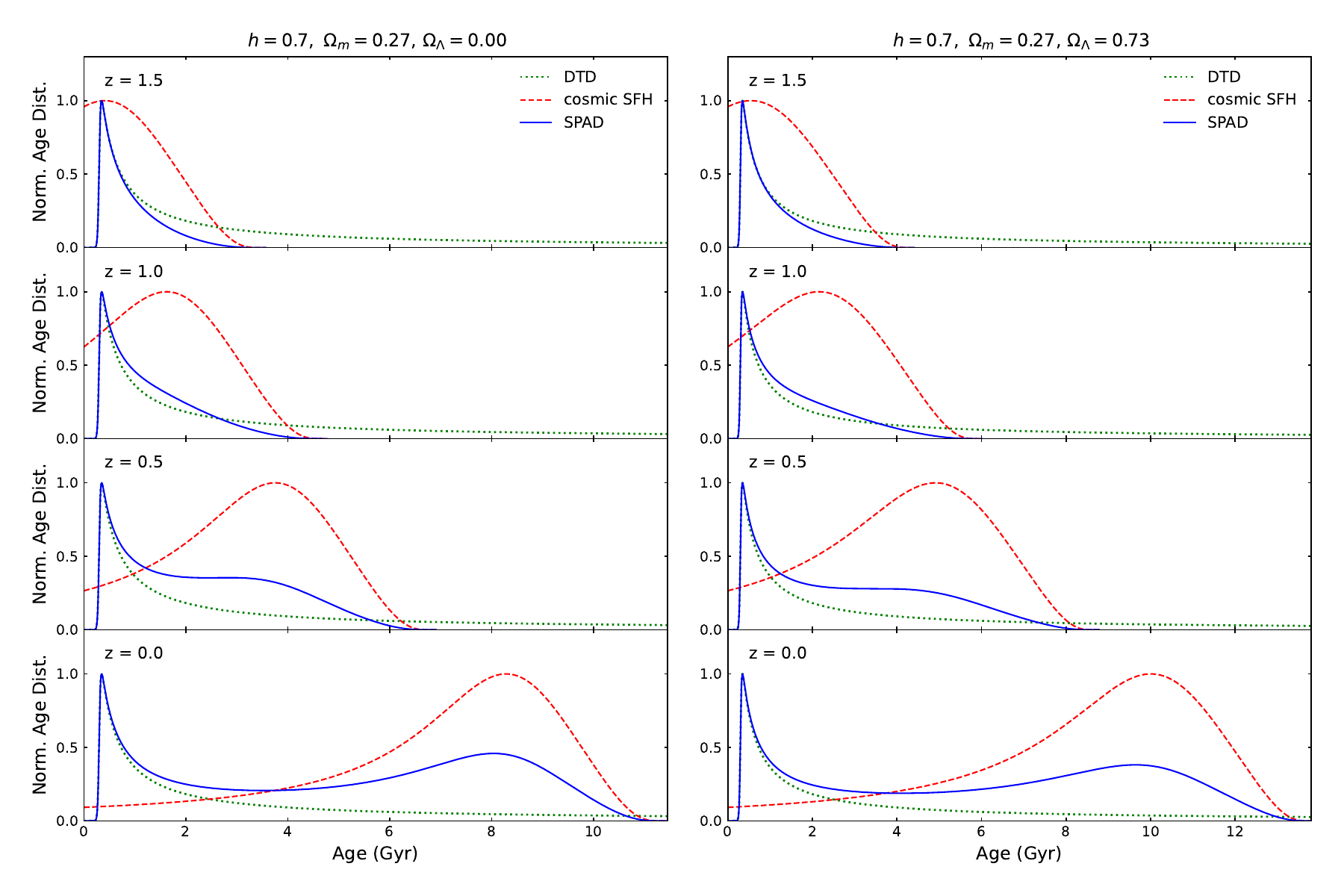}
\caption{ 
The SN Ia progenitor age distributions (SPADs) at $z = 0.0$, 0.5, 1.0, \& 1.5 under two different cosmological models ($\Omega_{\Lambda} = 0.00$ \& $\Omega_{\Lambda} = 0.73$). 
Following \citet{2014MNRAS.445.1898C} and \citet{2020ApJ...889....8K}, the SPAD at given redshift (the blue solid line) is derived by convolving the SN Ia delay time distribution (DTD, the green dotted line) with cosmic star formation history (SFH, the red dashed line).  
Note that the DTD has a peak at $\sim$0.3~Gyr but it also has a long tail towards older age, and since the cosmic SFH has a peak at $\sim$9~Gyr, the SPAD at $z = 0.0$ appears bimodal with a median age of $\sim$6~Gyr.
\label{f7}}
\end{figure*}

The population age dependence of {the zero-points of} the WLR and CLR is reminiscent of {the zero-point variation of the Cepheid period-luminosity relation} with age, and, as such, would be universal, and should have a critical impact on SN cosmology once and for all.
This result is quite robust to the choices of $\alpha$ and $\beta$, and of SN catalog.
For example, if we had adopted $\beta = 3.69$, as derived by \citet{2018ApJ...854...24K}, the difference between the two age subgroups is slightly larger ($0.177 \pm 0.041$~mag) at $x_1 = 0.0$.
If a larger value of $\alpha$ was adopted, $\alpha = 0.22$ as originally derived by \citet{2013ApJ...763...88C}, the difference becomes similarly larger ($0.184 \pm 0.036$~mag) at $c = 0.0$.
{The \citet{2019ApJ...874...32R} sample contains both spectroscopically and photometrically classified SNe Ia that passed cosmology cuts \citep{2013ApJ...763...88C} and additional cuts applied by them, but a similar $3.6\sigma$ difference is obtained even when only spectroscopically classified SNe Ia (77 SNe in \citealt{2014A&A...568A..22B} catalog in common with \citealt{2019ApJ...874...32R}) are used in the analysis.}
We have also tested this result with a mass-weighted age dataset by \citet{2011ApJ...740...92G} for a larger sample ($N = 206$) of host galaxies in a larger redshift range ($0.06 < z < 0.41$). 
This dataset is for the global ages (instead of local ages) of host galaxies based on an early version of the \citet{2010ApJ...712..833C} model, which would undermine the correlation, nonetheless, we still obtain a $3.1 \sigma$ difference between the young and old SN subsamples, partially confirming the universal nature of this stellar astrophysics effect.

What would be the origin of this shift in luminosity with progenitor age?
{We can speculate from Figure~\ref{f2}} that an increasing progenitor mass (with younger age) {would} produce a brighter SN with a broader light curve and bluer color.
{This imaginary vector in Figure~\ref{f2} would have a slope that is shallower than that of the WLR (or CLR) for old SNe (the red line), forming another WLR (or CLR) for younger SNe (the blue line). 
Therefore, younger SNe are intrinsically brighter, but, at given $x_1$ and $c$, they could be fainter.
This is the main point of our argument which would lead to the serious systematic bias in SN cosmology because the SN luminosity standardization is based on $x_1$ and $c$ {(see Section~\ref{s4} for the current practice of using a correction based on the host mass step instead of age)}.}
Theoretical models for type Ia SN are still incomplete, but a leading theory suggests that the WLR is mostly due to the asymmetry of ignition and detonation \citep{2009Natur.460..869K}, while other studies show that the SN peak luminosity increases with Ni mass formed in the explosion, which in turn increases with progenitor mass \citep{2007ApJ...662..487W, 2020arXiv201106513L}.
Therefore, taken together, the trend we can infer from {Figure~\ref{f2}} is theoretically plausible and qualitatively consistent with model predictions, while more detailed models are required for the quantitative comparison.
After the standardization (see {Figure~\ref{f4}}), which is nothing but the rotation of the WLR (or CLR) in {Figure~\ref{f2}} following the slope $\alpha$ (or $\beta$), SNe from younger progenitors are over-corrected and become relatively fainter.
The whole process of SN Ia luminosity standardization is illustrated again in {Figure~\ref{f5}}, based on the WLR and CLR, respectively, where the left panels are added to show the WLR and CLR before $x_1$ and $c$ corrections.
In the case of the WLR (upper panels), the split between the young and old progenitors is already visible in the left panel even before the $x_1$ and $c$ corrections, which becomes more pronounced after the $c$ correction in the middle panel.
In the case of the CLR (lower panels), however, the split is not shown in the left panel before $x_1$ and $c$ corrections.
It only becomes clear after the $x_1$ correction in the middle panel.
When $\mu_{\rm SN}$ is corrected for both $x_1$ and $c$ (after the luminosity standardization), SNe from younger progenitors become relatively fainter (right panels).
The values of HR in {Figures~\ref{f1} - \ref{f5}} would be fainter by 0.094~mag if they were calculated with respect to the non-$\Lambda$CDM model ($h = 0.70$, $\Omega_{M} = 0.27$, $\Omega_{\Lambda} = 0.00$).

\section{Comparison with cosmology sample}
\label{s3}

\begin{figure*}
\centering
\includegraphics[angle=0,scale=1.7]{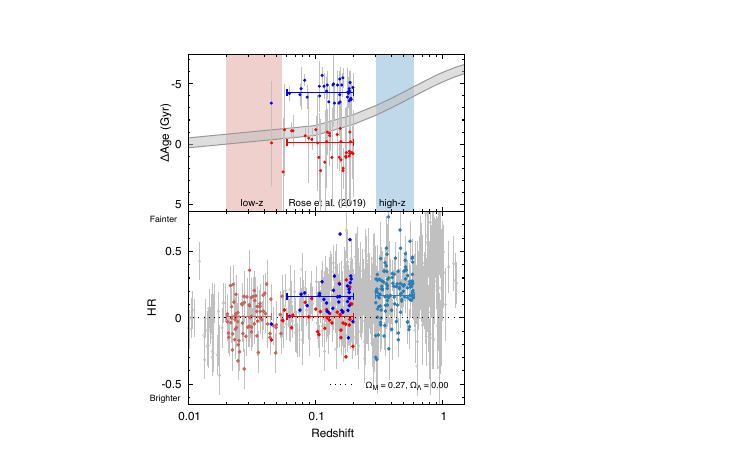}
\caption{
Comparison of \citet{2019ApJ...874...32R} sample with the cosmology sample (JLA dataset from \citealt{2014A&A...568A..22B}).
In the upper panel, the gray band is for the relative difference in average population age and its variation with redshift encompassing the two cosmological models of {Figure~\ref{f6}}, including the age uncertainty from $\pm 5 \%$ error in $H_0$.
The relative ages are with respect to $z = 0$ for the models and with respect to the old subsample for the \citet{2019ApJ...874...32R} sample.
{In the lower panel, the difference in HR between the high-$z$ and low-$z$ subsamples of the JLA data (cyan \& magenta circles colored solely because of their redshift) is fully consistent with that between the correspondingly young and old SNe (blue \& red circles) of \citet{2019ApJ...874...32R} sample at $z \sim 0.14$.
This illustrates that the observed dimming of SNe with redshift might not be caused by an accelerating universe, but rather by an evolution of the average progenitor age.}
As in the discovery paper \citep{1998AJ....116.1009R}, the HR's are calculated from the baseline model without dark energy ($h = 0.70$, $\Omega_{M} = 0.27$, $\Omega_{\Lambda} = 0.00$).
\label{f8}}
\end{figure*}

\begin{figure*}
\centering
\includegraphics[angle=-90,scale=0.85]{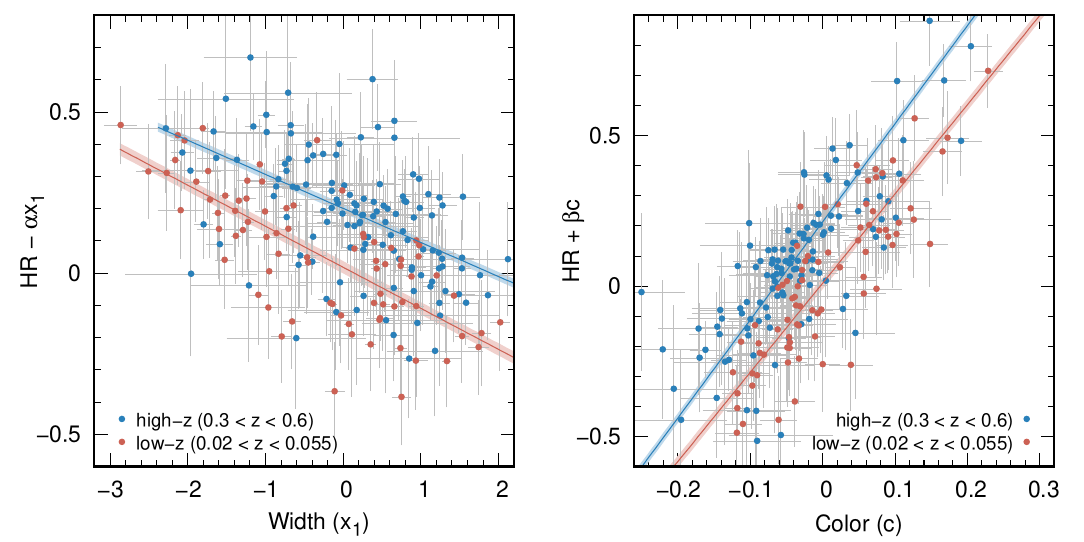}
\caption{
The WLR and CLR for the cosmology sample.
The shifts in the WLR and CLR between the high-$z$ and low-$z$ subsamples are similar to those between the young and old SNe of \citet{2019ApJ...874...32R} sample in {Figure~\ref{f2}} after an HR offset of 0.094~mag for the difference in the baseline model.
\label{f9}}
\end{figure*}

{As the redshift increases, the age of the universe, the maximum age possible for stellar populations, becomes younger. 
The average age of stellar populations at given redshift is therefore also getting younger with redshift, which is well supported by the observed population ages for galaxies at different redshifts (see Figure~\ref{f6}). 
As shown by \citet{2014MNRAS.445.1898C} and \citet{2020ApJ...889....8K}, which is illustrated in Figure~\ref{f7}, the SN progenitor age distribution (SPAD) at a given redshift can be obtained by convolving the {relatively} well-established {\citep{2013ApJ...770...57B, 2014ARA&A..52..415M}} cosmic star formation history (SFH), which is an ensemble average of SFHs of all galaxies, with the empirically derived SN Ia delay time distribution (DTD). 
These models naturally predict that the median age of SN progenitors gets younger with increasing redshift, following the average age of stellar populations, as the old populations no longer contribute to the SPAD at high-$z$ (see Figure~\ref{f6}). 
Within the redshift range most relevant to SN cosmology ($0 < z < 1$), we expect $\sim$$6$~Gyrs of variation in mean stellar population age, and a similar variation ($\sim$$5$~Gyrs) in median progenitor age (see Figure~\ref{f6} insets). 
As shown by \citet{2014MNRAS.445.1898C}, this variation in progenitor age is not strongly affected by current uncertainties in DTD. 
We believe, therefore, that this is a reasonable and probably the most appropriate approach we can take for now to estimate the SPAD and its evolution with redshift.}

The population age dependence of WLR and CLR discovered in the previous section is based on the \citet{2019ApJ...874...32R} sample at $z \sim 0.14$, which is also supported by the \citet{2011ApJ...740...92G} sample in $0.06 < z < 0.41$.
{Like the population age dependence of Cepheid} period-luminosity relation, which was originally discovered at the local universe but is valid at all distances, this effect should be universal.
Since the average age of stellar populations (SN progenitors) in host galaxies gets younger with redshift (see {Figure~\ref{f6}}), this progenitor age dependence of the WLR and CLR would naturally lead to the relative dimming of SNe with increasing redshift.
To investigate this effect more directly with the cosmology sample, in {Figure~\ref{f8}}, we compare the \citet{2019ApJ...874...32R} sample at $z \sim 0.14$ with the JLA dataset from \citet{2014A&A...568A..22B}.
Here, as in the discovery papers \citep{1998AJ....116.1009R, 2003LNP...598..195P}, the HR's are calculated from the baseline model without the dark energy ($h = 0.70$, $\Omega_{M} = 0.27$, $\Omega_{\Lambda} = 0.00$, $M_B=-19.08$) using the same parameters adopted for the \citet{2019ApJ...874...32R} sample ($\alpha = 0.16$, $\beta = 3.12$).
As compared in the upper panel of {Figure~\ref{f8}}, the high-$z$ subsample is chosen at $0.3 < z < 0.6$ so that the population age difference predicted between the low-$z$ ($0.02 < z < 0.055$) and high-$z$ subsamples would be roughly comparable to that between the old and young subsamples of \citet{2019ApJ...874...32R} at intermediate redshift.
The redshift range chosen for our high-$z$ subsample was also the redshift range for high-$z$ SNe of \citet[][their Figure 4]{1998AJ....116.1009R}, and, therefore, provides a fair comparison with the early analysis in a discovery paper.
Note that the variation of mean age with redshift predicted in {Figure~\ref{f6}} is confirmed by the observed data for the host and (presently) {non~Ia-host} galaxies at different redshift bins.

{In the lower panel of Figure~\ref{f8}, the observed difference in HR between the young and old subsamples of \citet{2019ApJ...874...32R} is $\sim$0.16~mag while the average ages of these two populations differ by $\sim$4~Gyr. 
We remark that, between $z = 0.0$ and $z \sim 0.5$, the average age of stellar population has changed by a similar amount. 
Between this redshift range, the average SNe Ia magnitude is getting $\sim$0.16~mag fainter than what would have been expected if the universe had no dark energy ($h=0.70$, $\Omega_m=0.27$, $\Omega_{\Lambda}=0.00$). 
In this context, we highlight that the origin of this magnitude difference observed between the low- and high-$z$ SNe Ia might not be caused by an accelerating universe, but rather by an evolution of the average progenitor age.}
Young SNe at any redshift, young SN subsample at $z \sim 0.14$ or high-$z$ sample as a whole, should be equally affected by the same progenitor age dependence of the WLR and CLR.
{Figure~\ref{f9}} confirms that the shifts in WLR and CLR between the high-$z$ and low-$z$ subsamples are similar to those between the young and old SNe of \citet{2019ApJ...874...32R} sample in {Figure~\ref{f2}} (after an HR offset of 0.094~mag for the difference in the baseline model), and, therefore, after the standardization, the high-$z$ SNe become similarly fainter.
The young SNe of \citet{2019ApJ...874...32R} sample are also similar to the high-$z$ cosmology sample in the distribution of light curve parameters $x_1$ and $c$.
In the $x_1$ versus $c$ plot, they all have the median values in the 4th quadrant (positive $x_1$ and negative $c$).
Specifically, the \citet{2019ApJ...874...32R} sample roughly predicts a mean variation in $x_1$ of $\sim$0.55 between $z \sim 0.14$ and 0.6 (average age $\sim$5.3 and 3.0 Gyr, respectively).
This is in reasonable agreement with the variation of $x_1$ ($\sim$$0.4 \pm 0.1$) in a similar redshift range reported by \citet[][see thier Figure 6]{2021A&A...649A..74N}.

In order to avoid the systematic bias from the progenitor age evolution with redshift, SNe from young and roughly coeval progenitors should be selected at all cosmological epochs.
The lower panel of {Figure~\ref{f8}} also offers a glimpse into this evolution-free cosmological test by using only the young SNe having more or less the same age at different redshifts (the blue and cyan circles), finding no dimming of high-$z$ SNe with respect to the equally young SNe at $z \sim 0.14$.
Therefore, we can confirm the self-consistency of our argument in several different ways.
This comparison and test indicate directly that an accelerating expansion of the universe, which was inferred from the observed dimming of SNe with redshift, may well be an artifact of over-correction in the SN luminosity standardization caused by the negligence of stellar astrophysics and stellar population effect.

\section{Discussion}
\label{s4}

{There are well-established steps in the standardized SN Ia luminosities with host galaxy mass or LsSFR \citep{2010ApJ...715..743K, 2010MNRAS.406..782S, 2013ApJ...770..108C, 2013A&A...560A..66R, 2020A&A...644A.176R}. 
It is important, therefore, to compare our result with these previous investigations. 
The most obvious difference is that the present work is based on directly measured population ages instead of employing host mass or LsSFR. 
As extensively discussed by \citet[][see their Table 7]{2020ApJ...889....8K}, other studies based on host mass, LsSFR, and host morphology, when transformed to the population age difference, are consistent with our result based on directly measured ages, suggesting that the root cause of these correlations is probably progenitor age. 
\citet{2020A&A...644A.176R} and \citet{2022A&A...657A..22B} also interpreted their results from LsSFR as progenitor age effects, while \citet{2010MNRAS.406..782S} suggested metallicity as a key physical variable driving the luminosity step, although they did not rule out the role of the progenitor age. 
Our main result is the {absolute magnitude differences} in the WLR and CLR between the two age subgroups. \citet{2010MNRAS.406..782S}, in particular, also showed that SNe Ia of the same light-curve shape and colour are fainter in less massive host galaxies and galaxies with high sSFR. 
Figure~10 of \citet{2010MNRAS.406..782S} shows a slight offset in the WLR between the two host mass subgroups, while Table~5 of \citet{2020A&A...644A.176R} presents a variation of standardization coefficients between the two subgroups based on LsSFR including a significant variation (0.13~mag) in the absolute magnitude.
Assuming host mass and LsSFR as rough proxies for age, their results are qualitatively consistent with our finding from directly measured population ages.}

{The \citet{2019ApJ...874...32R} sample employed in our analysis also provide other host properties (host mass, metallicity,  dust attenuation) and, therefore, we can directly compare the effects of these host properties with an age effect using the same sample in the WLR and CLR.} 
When the \citet{2019ApJ...874...32R} sample is {similarly} subdivided into two based on the host mass, unlike {Figure~\ref{f2}}, we find no clear {zero-point offset} between the two subgroups on the WLR and CLR.
The difference in {absolute magnitude} of the two host mass subgroups is at most $0.055 \pm 0.044$~mag ($1.3\sigma$), only one-third of the difference found in {Figure~\ref{f2}} between the two age subgroups.
Even if the mass step correction \citep{2014A&A...568A..22B} is added in Equation~(\ref{e1}), the magnitude difference between the two age subgroups in {Figure~\ref{f2}} is reduced by only 8\% (0.013~mag).\footnote{ This indicates that the use of the mass step, instead of age, can correct only a small fraction of the age effect.
In particular, unlike age, the variation in host mass is negligible within the redshift range most relevant to SN cosmology ($z < 1.0$), and, therefore, the current practice of using a correction based on the mass step cannot correct for the luminosity evolution with redshift originated from the age variation.}
This practically confirms that the apparent correlation with host mass is not directly originated from host mass, but is rather a reflection of a mild relationship between population age and host mass among galaxies \citep{2010MNRAS.404.1775T, 2016ApJS..223....7K}.
{Figure~8 of \citet{2005MNRAS.362...41G}, in particular, shows a nonlinear relation between galaxy mass and population age \citep[see also][]{2009MNRAS.397.1776F}, which could reproduce the mass step from a linear relation between HR and age.} 
Figure~7 of \citet{2014MNRAS.445.1898C} {also} explains how the host mass step is driven by progenitor age.
A similar test for the {LsSFR} would be impracticable, because our analysis on the GALEX UV color-magnitude diagram shows that most (95\%) galaxies in the \citet{2019ApJ...874...32R} sample would be classified as star-forming (on-going or recent SF) galaxies of \citet{2015ApJ...802...20R, 2020A&A...644A.176R}.
However, we note that, while there is no apparent correlation between {LsSFR} and HR among star-forming host galaxy subsample of \citet{2020A&A...644A.176R}, there is a strong correlation between population age and HR among similarly star-forming galaxies of \citet{2019ApJ...874...32R}.
This, together with the progenitor age dependence of WLR in the same sample galaxies, suggests that the progenitor age is the root cause of the {LsSFR}-HR correlation as well {as has been interpreted by \citet{2020A&A...644A.176R}}.

Recent studies \citep{2021ApJ...909...26B, 2021ApJ...909...28R} suggest dust attenuation and metallicicty may derive the intrinsic scatter of SN Ia HRs after standardization.
Using the same \citet{2019ApJ...874...32R} dataset, we have tested these suggestions by subdividing the sample into two based on the dust attenuation parameter $\tau_2$ and the metallicity index of \citet{2019ApJ...874...32R}.
Similar to our analysis for population age in {Figure~\ref{f2}}, each subgroup had $\sim$36 SNe with a gray zone between.
When measured at $x_1 = 0.0$, we find only $0.050 \pm 0.041$~mag ($1.2 \sigma$) and $0.044 \pm 0.045$~mag ($0.98 \sigma$) differences in the WLR between the two subgroups for the dust and metallicity, respectively.
This result is not sensitive to the choice in the split between the two subgroups.
These differences are substantially smaller and not statistically significant, suggsting dust and metallicity are not likely the root cause of the host property dependence of HR.
\citet{2020ApJ...889....8K} also find no correlation with metallicity from their high-quality spectroscopic sample of early-type host galaxies.
The young and old SN subsamples of \citet{2019ApJ...874...32R} in {Figure~\ref{f8}} directly show that most of the scatter in HR at a given redshift is due to population age, although we cannot rule out the possibility that dust and metallicity may play a secondary role.

In the cosmological application of this result, the most important information to recall is the population age distribution of host galaxies at low-$z$, for which \citet{2019ApJ...874...32R} and \citet{2020ApJ...889....8K} all show clearly that both young and old populations produce SNe Ia with the average age of $\sim$7~Gyr at the local universe.
This agrees well with the model predictions in {Figures~\ref{f6} and \ref{f7}}.
The same models also predict that the median age of SN progenitors varies gradually with redshift, similarly to the redshift variation of the average age of stellar populations (see {Figure~\ref{f6}} insets).
The cosmological analysis in the Hubble residual diagram is usually based on the mean or median value of HR at a given redshift, and, therefore, the use of the mean or median age would be the most reasonable choice in our analysis.
It is unavoidable to conclude from this reasoning that the population age dependence of the WLR and CLR discovered in this paper would lead to a significant luminosity evolution with redshift (look-back time) in SN cosmology.
This possibility was also clearly pointed out and extensively discussed by investigators of the discovery papers and of the recent literature \citep{1998ApJ...507...46S, 1998AJ....116.1009R, 1999ApJ...517..565P, 2019NatAs...3..706J}.

\begin{table}
\begin{center}
\caption{\label{t1} Luminosity evolution with redshift}
\begin{tabular}{cccccc}
\hline
 & \multicolumn{2}{c}{$\Omega_{\Lambda}=0.73$}&& \multicolumn{2}{c}{$\Omega_{\Lambda}=0.00$}\\
\cline{2-3} \cline{5-6}
$z$ & $\Delta {\rm age} $ (Gyr) & $\Delta {\rm mag}$ && $\Delta {\rm age}$ (Gyr) & $\Delta {\rm mag}$ \\
\hline
0.00 & 0.00 & 0.000 && 0.00 & 0.000 \\
0.10 & 1.11 & 0.044 && 1.07 & 0.042 \\
0.20 & 2.07 & 0.082 && 1.94 & 0.077 \\
0.30 & 2.89 & 0.115 && 2.65 & 0.106 \\
0.40 & 3.60 & 0.144 && 3.24 & 0.129 \\
0.50 & 4.21 & 0.168 && 3.73 & 0.149 \\
0.60 & 4.73 & 0.189 && 4.14 & 0.165 \\
0.70 & 5.17 & 0.206 && 4.49 & 0.179 \\
0.80 & 5.55 & 0.222 && 4.78 & 0.191 \\
0.90 & 5.88 & 0.235 && 5.03 & 0.201 \\
1.00 & 6.16 & 0.246 && 5.24 & 0.209 \\
1.10 & 6.40 & 0.256 && 5.42 & 0.216 \\
1.20 & 6.60 & 0.264 && 5.57 & 0.222 \\
1.30 & 6.77 & 0.270 && 5.70 & 0.228 \\
1.40 & 6.92 & 0.276 && 5.81 & 0.232 \\
1.50 & 7.04 & 0.281 && 5.90 & 0.236 \\
\hline
\multicolumn{6}{l}{Note. $\Omega_M=0.27$ and $h=0.70$ for all cases.}
\end{tabular}
\end{center}
\end{table}

\begin{figure*}
\centering
\includegraphics[angle=0,scale=0.6]{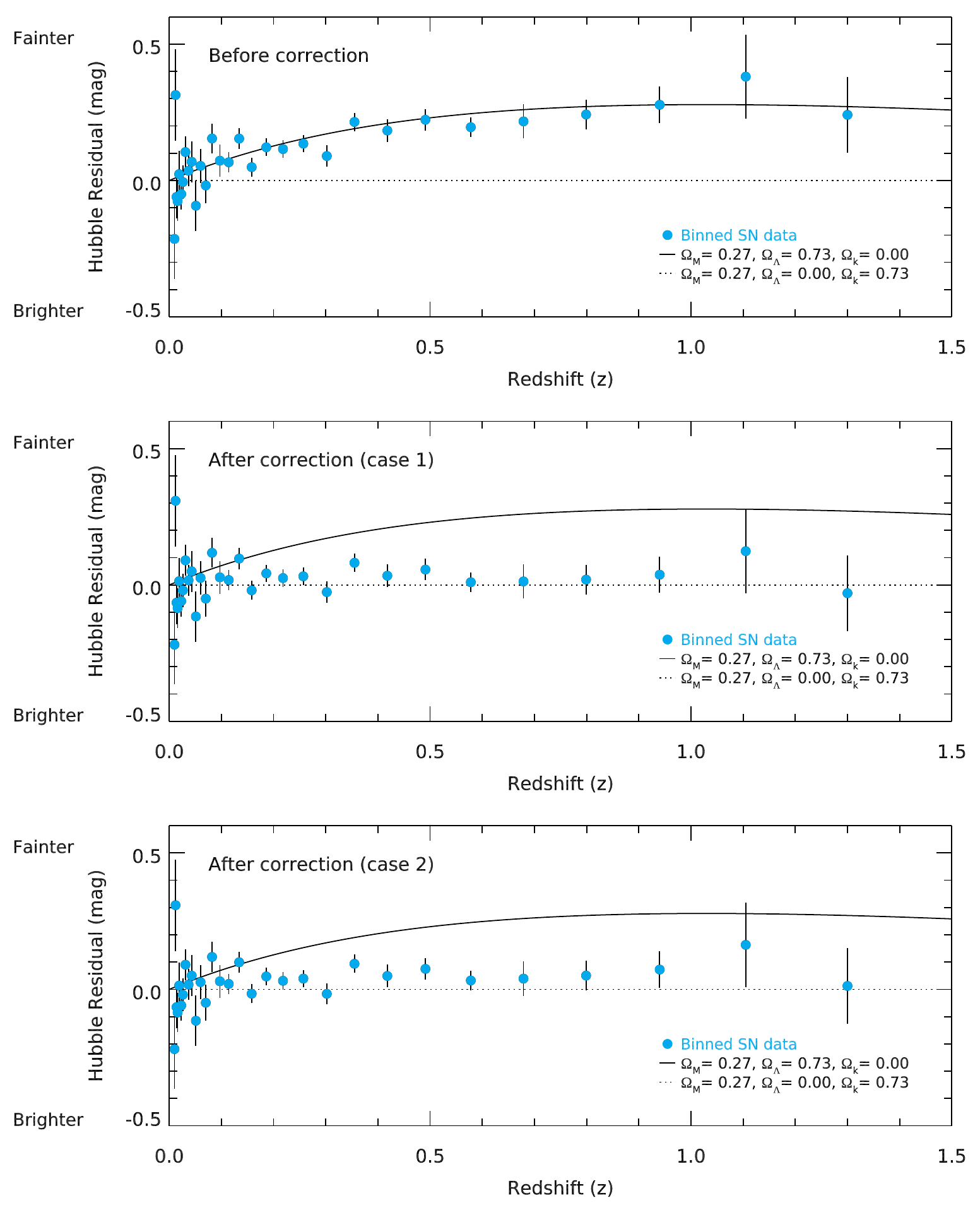}
\caption{
The residual Hubble diagram before and after the correction for the luminosity evolution.
The corrections for the luminosity evolution (Table 1) are made to the observational data \citep{2014A&A...568A..22B} using the $\Delta {\rm HR}/ \Delta {\rm age}$ slope ($-0.040$~mag/Gyr) and the redshift evolution of the mean population age in the right (case 1) and left (case 2) panels of {Figure~\ref{f6}}.
After the correction, there is little evidence left for an accelerating universe.
\label{f10}}
\end{figure*}

\begin{figure}
\centering
\includegraphics[angle=0,scale=0.33]{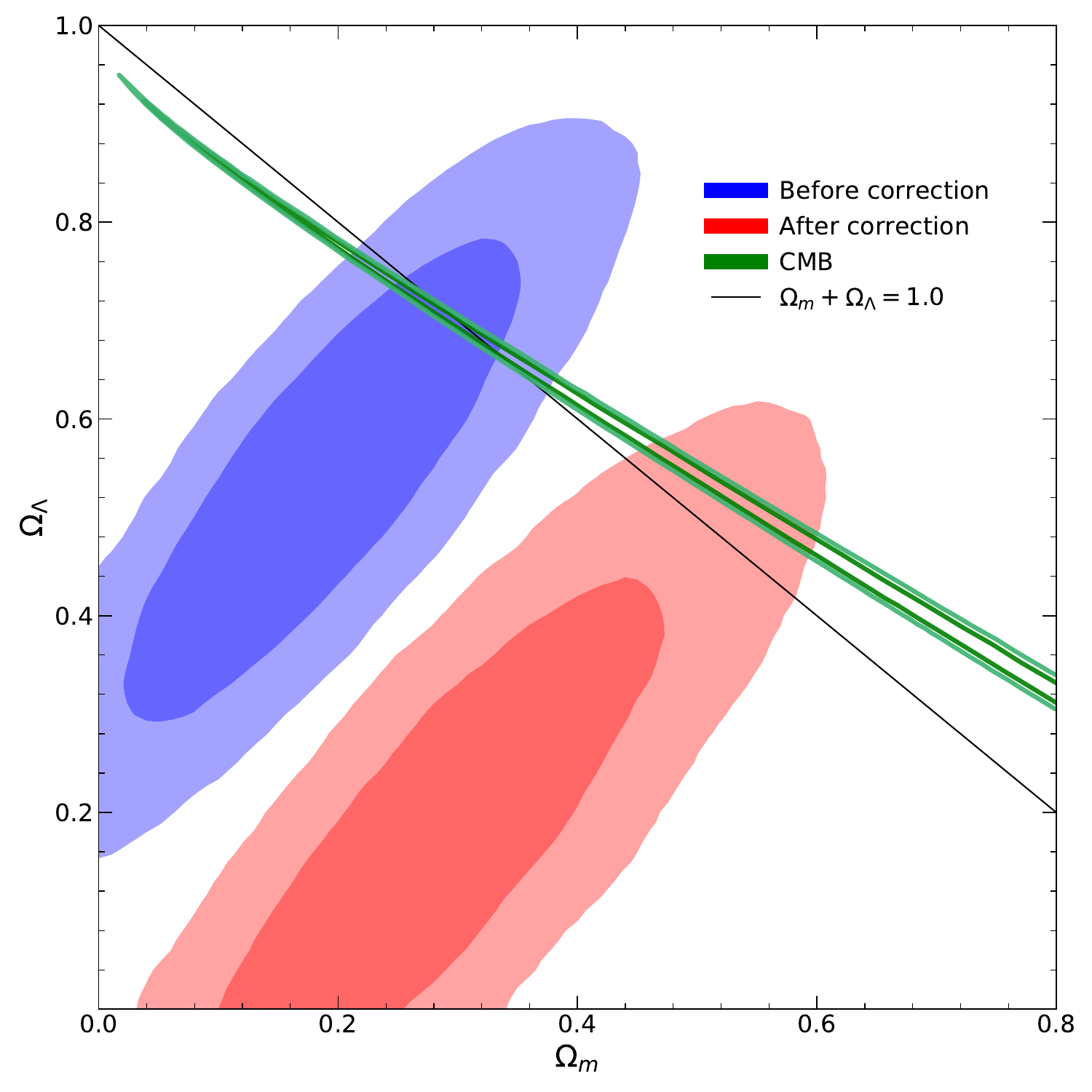}
\caption{
The effect of our age bias correction on the confidence contours (68\% and 95\%) for the $\Omega_m$ and $\Omega_{\Lambda}$ cosmological parameters for the SN dataset and the o-$\Lambda$CDM model of \citet{2014A&A...568A..22B}. 
After the correction, the SNe would show a discordance with CMB. 
The purpose of this comparison is only to illustrate, in a relative sense, the possible impact predicted from our findings.
For this diagram, we have followed the procedures in Sections 5-7 and Figure~15 of \citet{2014A&A...568A..22B}, together with the age bias correction self-consistently determined with the cosmological fit.
\label{f11}}
\end{figure}

In order to illustrate the level of significance of the luminosity evolution predicted from our finding, it is heuristically useful to see what they would have found if, at the discovery time, they knew the $\Delta {\rm HR} / \Delta{\rm age}$ slope {inferred from Figure~\ref{f2} of} this paper. 
{The value of this slope ($\Delta{\rm HR}/\Delta{\rm age} = -0.040$~mag/Gyr) is adopted from the magnitude offset in Figure~\ref{f2} assuming a general slope of a linear correlation. 
Note that it is similar to those reported by \citet{2020ApJ...889....8K} from early-type host galaxies ($\Delta{\rm HR}/\Delta{\rm age} = -0.051 \pm 0.022$~mag/Gyr) and by \citet{2020ApJ...903...22L} from the \citet{2019ApJ...874...32R} sample comprising all morphological types ($\Delta{\rm HR}/\Delta{\rm age} = -0.057 \pm 0.016$~mag/Gyr). 
The statistical significance of the slope obtained by \citet{2020ApJ...903...22L} is confirmed by \citet{2021MNRAS.503L..33Z} with a somewhat shallower slope ($\Delta{\rm HR}/\Delta{\rm age} = -0.035 \pm 0.007$~mag/Gyr). 
A straight mean of these three values is $-0.048$~mag/Gyr, comparable to the value adopted in this paper. 
Furthermore, \citet{2020ApJ...903...22L} noted that the potential effect of a nonlinearity is not significant in their age-HR correlation, in the age range ($t < 8$~Gyr) most relevant to SN cosmology, in particular. 
This should be further confirmed, however, from a larger sample of host galaxies at different redshift bins.}
Table~\ref{t1} lists our predictions of the magnitude corrections for the SN Ia luminosity evolution with redshift under two different assumptions for the cosmological model in {Figure~\ref{f6}}.
The corrections are made based on the differences in population age with respect to $z=0.0$ ({Figure~\ref{f6}} insets) with a slope $\Delta {\rm HR} / \Delta {\rm age} = -0.040$~mag/Gyr.

{Our approach differs from the early pioneering method employing a simple two-component (`prompt and delayed') model of SNe Ia \citep{2006ApJ...648..868S, 2012PASA...29..447M}. 
Instead of just adopting a variation in the relative fraction of the two components with redshift, our model is based on the cosmic SFH and SPAD and their evolution with redshift in Figures~\ref{f6} and \ref{f7}.
The SPAD at $z < 0.5$ shows a bimodal distribution, which would qualitatively correspond to the prompt (young) and delayed (old) components. 
As in the previous models, the peak age of the young (prompt) population in the SPAD remains almost the same as a function of redshift. 
But the maximum age for the old (delayed) population varies significantly with redshift (see the red line in Figure~\ref{f6}). 
In addition, the SPAD no longer shows a bimodal distribution at $z > 0.5$ with no clear separation between the young and old populations. 
It only shows a single peak at a young age with a long tail towards old ages. 
Therefore, both the median age of SN progenitors and the average age of stellar populations show similar differences at high z with respect to $z = 0$ (see the insets in Figure~\ref{f6}). 
Our approach is also different from the model taken in \citet{2020A&A...644A.176R} and \citet{2021A&A...649A..74N}. 
Instead of adopting a fixed value ($0.16$~mag) for the SN Ia magnitude difference between the star-forming (young) and passive (old) environments, we have used the slope, $\Delta {\rm HR}/\Delta {\rm age}$, and the relative difference in age with respect to $z = 0$ to estimate the magnitude correction at a given redshift. 
Furthermore, as described above, instead of the redshift evolution of the fraction of the passive (old) and star-forming (young) host galaxies, we have directly employed the cosmic SFH and SPAD and their evolution with redshift. 
Therefore, the age bias correction based on directly measured ages in this paper makes a great difference from previous investigations based on host mass or LsSFR.
Although \citet{2020A&A...644A.176R} interpreted their result from LsSFR as a progenitor age effect, previous investigations considered only the redshift evolution of host mass or LsSFR \citep{2018ApJ...859..101S, 2020A&A...644A.176R} instead of the age variation with redshift. 
Unlike age, the variations of these host properties are either negligible or relatively small within the redshift range most relevant to SN cosmology ($0 < z < 1$), yielding only an insignificant or limited impact on cosmology.}

{Figure~\ref{f10}} shows the effects of {our} corrections for the SN Ia luminosity evolution in the residual Hubble diagram.
{We want to stress here that the purpose of this comparison is only to illustrate, in the relative sense, the possible impact of the luminosity evolution predicted from our finding, and is not to derive or suggest the cosmological model that best matches the data points.}
The SN data over-plotted are from the binned values of \citet{2014A&A...568A..22B}.
Case 1 is for {the redshift - $\Delta {\rm age}$ relation} obtained with $\Omega_{\Lambda} = 0.73$, while case 2 is for that with $\Omega_{\Lambda} = 0.00$.
The difference between the two cases is small, if not negligible, which illustrates that our main conclusion below is valid regardless of the pre-assumed cosmological parameters used in {the redshift - $\Delta {\rm age}$ conversion.}
As expected from {Figure~\ref{f8}}, {Figure~\ref{f10}} shows clearly that, when the luminosity evolution is properly corrected, the data points are distributed close to the non-accelerating model (the dotted line), rather than an accelerating universe with $\Omega_{\Lambda} = 0.73$ (the solid line).
Specifically, for $z > 0.3$, each of the data points is, on average, $\sim$$3.7 \sigma$ away from the accelerating universe model, while it is within $\sim$$0.8 \sigma$ from the non-accelerating model.
In this redshift range, a $\chi^2$ test also confirms that the accelerating universe model does not fit the data points with a large value for the reduced $\chi^2_\nu$ (14.5 - 18.1), while the reference model (non-accelerating) yields a reduced $\chi^2_\nu$ value close to 1 (1.3 - 2.2).
If we had used progenitor ages, instead of population ages, the deviation from the accelerating universe model would be even larger or similar because the $\Delta {\rm HR}/\Delta {\rm age}$ slope would be steeper, as described in Section~\ref{s2}, while the progenitor age variation with redshift is somewhat smaller (see {Figure~\ref{f6}} insets).
Therefore, taken at face values, there is little evidence left for an accelerating universe when the progenitor age dependence of the SN luminosity standardization is properly taken into account.
It appears that {most of} the observed dimming of SNe with redshift may not be due to the cosmological effect, but {may well be} due to the stellar astrophysics and stellar population effect.

The SN cosmology has long been considered {the first and most direct evidence for an accelerating universe,} and, as such, it forms one of the cornerstones of the concordance model together with CMB and {BAO} \citep{2008ARA&A..46..385F, 2013PhR...530...87W}.
This finding, therefore, poses {some serious concerns} to one of these cornerstones (SNe Ia), but {would be} in discordance with other cosmological measures from CMB and BAO \citep{2020A&A...641A...6P, 2021PhRvD.103h3533A} within the Friedmann-Lemaitre-Robertson-Walker model {(see Figure~\ref{f11})}.
{Even after the well-established results from CMB, we believe that it is still important to check the potential systematic bias in SN cosmology, as it provides the most direct evidence for an accelerating universe. 
This should be performed in a completely independent manner from other probes, because, otherwise, it will be very difficult to discover this bias even if any. 
This paper is part of this endeavor.}
While the present result {is interesting}, to put this result on a firmer refined basis, reliable mass-weighted ages (\`a la \citealt{2019ApJ...874...32R}) for stellar populations near the SN sites in host galaxies would be required for a larger sample of host galaxies at different redshift bins.
If the present result is further supported by these studies, an important avenue of future investigations would be to see how this result from SNe can be reconciled with other cosmological probes.

\section*{Acknowledgements}
We thank David Weinberg, James Jee, Mark Sullivan, Robert Zinn, Stan Woosley, William Forrest for their comments on the early draft of this paper.
{We also thank the referee for a number of helpful comments and suggestions which led to several improvements in the manuscript.}
Support for this work was provided by the National Research Foundation of Korea (2022R1A2C3002992, 2022R1A6A1A03053472).

\section*{Data Availability}

There are no new data associated with this article.



\bibliographystyle{mnras}

\begin{thebibliography}{}
\expandafter\ifx\csname natexlab\endcsname\relax\def\natexlab#1{#1}\fi

\bibitem[Alam et al.(2021)]{2021PhRvD.103h3533A} Alam, S., Aubert, M., Avila, S., et al.\ 2021, \prd, 103, 083533. doi:10.1103/PhysRevD.103.083533

\bibitem[Astier et al.(2006)]{2006A&A...447...31A} Astier, P., Guy, J., Regnault, N., et al.\ 2006, \aap, 447, 31. doi:10.1051/0004-6361:20054185

\bibitem[Baade(1956)]{1956PASP...68....5B} Baade, W.\ 1956, \pasp, 68, 5. doi:10.1086/126870

\bibitem[Behroozi et al.(2013)]{2013ApJ...770...57B} Behroozi, P.~S., Wechsler, R.~H., \& Conroy, C.\ 2013, \apj, 770, 57. doi:10.1088/0004-637X/770/1/57

\bibitem[Betoule et al.(2014)]{2014A&A...568A..22B} Betoule, M., Kessler, R., Guy, J., et al.\ 2014, \aap, 568, A22

\bibitem[Briday et al.(2022)]{2022A&A...657A..22B} Briday, M., Rigault, M., Graziani, R., et al.\ 2022, \aap, 657, A22. doi:10.1051/0004-6361/202141160

\bibitem[Brout \& Scolnic(2021)]{2021ApJ...909...26B} Brout, D. \& Scolnic, D.\ 2021, \apj, 909, 26. doi:10.3847/1538-4357/abd69b

\bibitem[Campbell et al.(2013)]{2013ApJ...763...88C} Campbell, H., D'Andrea, C.~B., Nichol, R.~C., et al.\ 2013, \apj, 763, 88

\bibitem[Childress et al.(2013)]{2013ApJ...770..108C} Childress, M., Aldering, G., Antilogus, P., et al.\ 2013, \apj, 770, 108. doi:10.1088/0004-637X/770/2/108

\bibitem[Childress et al.(2014)]{2014MNRAS.445.1898C} Childress, M.~J., Wolf, C., \& Zahid, H.~J.\ 2014, \mnras, 445, 1898

\bibitem[Choi et al.(2014)]{2014ApJ...792...95C} Choi, J., Conroy, C., Moustakas, J., et al.\ 2014, \apj, 792, 95. doi:10.1088/0004-637X/792/2/95

\bibitem[Colin et al.(2019)]{2019A&A...631L..13C} Colin, J., Mohayaee, R., Rameez, M., et al.\ 2019, \aap, 631, L13. doi:10.1051/0004-6361/201936373

\bibitem[Conroy \& Gunn(2010)]{2010ApJ...712..833C} Conroy, C., \& Gunn, J.~E.\ 2010, \apj, 712, 833

\bibitem[Daly et al.(2008)]{2008ApJ...677....1D} Daly, R.~A., Djorgovski, S.~G., Freeman, K.~A., et al.\ 2008, \apj, 677, 1. doi:10.1086/528837

\bibitem[Drell et al.(2000)]{2000ApJ...530..593D} Drell, P.~S., Loredo, T.~J., \& Wasserman, I.\ 2000, \apj, 530, 593. doi:10.1086/308393

\bibitem[Fontanot et al.(2009)]{2009MNRAS.397.1776F} Fontanot, F., De Lucia, G., Monaco, P., et al.\ 2009, \mnras, 397, 1776. doi:10.1111/j.1365-2966.2009.15058.x

\bibitem[Frieman et al.(2008)]{2008ARA&A..46..385F} Frieman, J.~A., Turner, M.~S., \& Huterer, D.\ 2
008, \araa, 46, 385. doi:10.1146/annurev.astro.46.060407.145243

\bibitem[Fumagalli et al.(2016)]{2016ApJ...822....1F} Fumagalli, M., Franx, M., van Dokkum, P., et al.\ 2016, \apj, 822, 1. doi:10.3847/0004-637X/822/1/1

\bibitem[Gallazzi et al.(2005)]{2005MNRAS.362...41G} Gallazzi, A., Charlot, S., Brinchmann, J., et al.\ 2005, \mnras, 362, 41. doi:10.1111/j.1365-2966.2005.09321.x

\bibitem[Gupta et al.(2011)]{2011ApJ...740...92G} Gupta, R.~R., D'Andrea, C.~B., Sako, M., et al.\ 2011, \apj, 740, 92

\bibitem[Guy et al.(2007)]{2007A&A...466...11G} Guy, J., Astier, P., Baumont, S., et al.\ 2007, \aap, 466, 11. doi:10.1051/0004-6361:20066930

\bibitem[Jha et al.(2019)]{2019NatAs...3..706J} Jha, S.~W., Maguire, K., \& Sullivan, M.\ 2019, Nature Astronomy, 3, 706. doi:10.1038/s41550-019-0858-0

\bibitem[Johansson et al.(2013)]{2013MNRAS.435.1680J} Johansson, J., Thomas, D., Pforr, J., et al.\ 2013, \mnras, 435, 1680. doi:10.1093/mnras/stt1408

\bibitem[Jones et al.(2018)]{2018ApJ...867..108J} Jones, D.~O., Riess, A.~G., Scolnic, D.~M., et al.\ 2018, \apj, 867, 108

\bibitem[Kang et al.(2016)]{2016ApJS..223....7K} Kang, Y., Kim, Y.-L., Lim, D., et al.\ 2016, \apjs, 223, 7. doi:10.3847/0067-0049/223/1/7

\bibitem[Kang et al.(2020)]{2020ApJ...889....8K} Kang, Y., Lee, Y.-W., Kim, Y.-L., et al.\ 2020, \apj, 889, 8

\bibitem[Kasen et al.(2009)]{2009Natur.460..869K} Kasen, D., R{\"o}pke, F.~K., \& Woosley, S.~E.\ 2009, \nat, 460, 869. doi:10.1038/nature08256

\bibitem[Kelly(2007)]{2007ApJ...665.1489K} Kelly, B.~C.\ 2007, \apj, 665, 1489. doi:10.1086/519947

\bibitem[Kelly et al.(2010)]{2010ApJ...715..743K} Kelly, P.~L., Hicken, M., Burke, D.~L., et al.\ 2010, \apj, 715, 743

\bibitem[Kim et al.(2018)]{2018ApJ...854...24K} Kim, Y.-L., Smith, M., Sullivan, M., et al.\ 2018, \apj, 854, 24


\bibitem[Lee et al.(2020)]{2020ApJ...903...22L} Lee, Y.-W., Chung, C., Kang, Y., et al.\ 2020, \apj, 903, 22. doi:10.3847/1538-4357/abb3c6

\bibitem[Leung et al.(2020)]{2020arXiv201106513L} Leung, S.-C., Diehl, R., Nomoto, K., et al.\ 2020, arXiv:2011.06513

\bibitem[Linden et al.(2009)]{2009A&A...506.1095L} Linden, S., Virey, J.-M., \& Tilquin, A.\ 2009, \aap, 506, 1095. doi:10.1051/0004-6361/200912811

\bibitem[Madau \& Dickinson(2014)]{2014ARA&A..52..415M} Madau, P. \& Dickinson, M.\ 2014, \araa, 52, 415. doi:10.1146/annurev-astro-081811-125615

\bibitem[Maoz \& Mannucci(2012)]{2012PASA...29..447M} Maoz, D. \& Mannucci, F.\ 2012, \pasa, 29, 447. doi:10.1071/AS11052

\bibitem[Nicolas et al.(2021)]{2021A&A...649A..74N} Nicolas, N., Rigault, M., Copin, Y., et al.\ 2021, \aap, 649, A74. doi:10.1051/0004-6361/202038447


\bibitem[Phillips(1993)]{1993ApJ...413L.105P} Phillips, M.~M.\ 1993, \apjl, 413, L105. doi:10.1086/186970

\bibitem[Perlmutter et al.(1999)]{1999ApJ...517..565P} Perlmutter, S., Aldering, G., Goldhaber, G., et al.\ 1999, \apj, 517, 565

\bibitem[Perlmutter \& Schmidt(2003)]{2003LNP...598..195P} Perlmutter, S. \& Schmidt, B.~P.\ 2003, Supernovae and Gamma-Ray Bursters, 195. doi:10.1007/3-540-45863-8\_11

\bibitem[Planck Collaboration et al.(2020)]{2020A&A...641A...6P} Planck Collaboration, Aghanim, N., Akrami, Y., et al.\ 2020, \aap, 641, A6. doi:10.1051/0004-6361/201833910

\bibitem[Riess et al.(1998)]{1998AJ....116.1009R} Riess, A.~G., Filippenko, A.~V., Challis, P., et al.\ 1998, \aj, 116, 1009

\bibitem[Rigault et al.(2015)]{2015ApJ...802...20R} Rigault, M., Aldering, G., Kowalski, M., et al.\ 2015, \apj, 802, 20

\bibitem[Rigault et al.(2020)]{2020A&A...644A.176R} Rigault, M., Brinnel, V., Aldering, G., et al.\ 2020, \aap, 644, A176. doi:10.1051/0004-6361/201730404


\bibitem[Rigault et al.(2013)]{2013A&A...560A..66R} Rigault, M., Copin, Y., Aldering, G., et al.\ 2013, \aap, 560, A66. doi:10.1051/0004-6361/201322104

\bibitem[Rose et al.(2019)]{2019ApJ...874...32R} Rose, B.~M., Garnavich, P.~M., \& Berg, M.~A.\ 2019, \apj, 874, 32

\bibitem[Rose et al.(2020)]{2020ApJ...896L...4R} Rose, B.~M., Rubin, D., Cikota, A., et al.\ 2020, \apjl, 896, L4

\bibitem[Rose et al.(2021)]{2021ApJ...909...28R} Rose, B.~M., Rubin, D., Strolger, L., et al.\ 2021, \apj, 909, 28. doi:10.3847/1538-4357/abd550


\bibitem[Schiavon et al.(2006)]{2006ApJ...651L..93S} Schiavon, R.~P., Faber, S.~M., Konidaris, N., et al.\ 2006, \apjl, 651, L93. doi:10.1086/509074

\bibitem[Schmidt et al.(1998)]{1998ApJ...507...46S} Schmidt, B.~P., Suntzeff, N.~B., Phillips, M.~M., et al.\ 1998, \apj, 507, 46

\bibitem[Scolnic et al.(2018)]{2018ApJ...859..101S} Scolnic, D.~M., Jones, D.~O., Rest, A., et al.\ 2018, \apj, 859, 101. doi:10.3847/1538-4357/aab9bb

\bibitem[Sullivan et al.(2010)]{2010MNRAS.406..782S} Sullivan, M., Conley, A., Howell, D.~A., et al.\ 2010, \mnras, 406, 782

\bibitem[Sullivan et al.(2006)]{2006ApJ...648..868S} Sullivan, M., Le Borgne, D., Pritchet, C.~J., et al.\ 2006, \apj, 648, 868. doi:10.1086/506137

\bibitem[Thomas et al.(2010)]{2010MNRAS.404.1775T} Thomas, D., Maraston, C., Schawinski, K., et al.\ 2010, \mnras, 404, 1775. doi:10.1111/j.1365-2966.2010.16427.x

\bibitem[Tinsley(1968)]{1968ApJ...151..547T} Tinsley, B.~M.\ 1968, \apj, 151, 547. doi:10.1086/149455

\bibitem[Tripp(1998)]{1998A&A...331..815T} Tripp, R.\ 1998, \aap, 331, 815

\bibitem[Tutusaus et al.(2017)]{2017A&A...602A..73T} Tutusaus, I., Lamine, B., Dupays, A., et al.\ 2017, \aap, 602, A73. doi:10.1051/0004-6361/201630289

\bibitem[Wallerstein(2002)]{2002PASP..114..689W} Wallerstein, G.\ 2002, \pasp, 114, 689. doi:10.1086/341698

\bibitem[Weinberg et al.(2013)]{2013PhR...530...87W} Weinberg, D.~H., Mortonson, M.~J., Eisenstein, D.~J., et al.\ 2013, \physrep, 530, 87. doi:10.1016/j.physrep.2013.05.001

\bibitem[Woosley et al.(2007)]{2007ApJ...662..487W} Woosley, S.~E., Kasen, D., Blinnikov, S., et al.\ 2007, \apj, 662, 487. doi:10.1086/513732

\bibitem[Zhang et al.(2021)]{2021MNRAS.503L..33Z} Zhang, K.~D., Murakami, Y.~S., Stahl, B.~E., et al.\ 2021, \mnras, 503, L33. doi:10.1093/mnrasl/slab020

\end{thebibliography}









\bsp	
\label{lastpage}
\end{document}